\documentclass[aps,pre,twocolumn,superscriptaddress]{revtex4-1}
\usepackage{epsfig,amsmath,amssymb,amsfonts,physics,graphicx,color,calc,epstopdf}
\usepackage[ruled,vlined]{algorithm2e}

\usepackage{xcolor,hyperref}
\hypersetup{
   colorlinks,
   linkcolor={blue!50!black},
   citecolor={blue!50!black},
   urlcolor={blue!80!black}
}

 \let\b=\beta  \let\d=\delta
\let\e=\varepsilon  \let\h=\eta \let\k=\kappa
\let\l=\lambda  \let\n=\nu  
   \let\c=\chi
   
\let\D=\Delta \let\L=\Lambda  
   
\let\ee=\epsilon \let\r=\rho \let\th=\theta \let\io=\infty

\def\MM{{\cal M}} 
\def\FF{{\cal F}} 
\def\NN{{\cal N}} \def\BB{{\cal B}}\def\II{{\cal I}}
  
\def\DD{{\cal D}}\def\AA{{\cal A}}

\def\Im{{\rm Im}\,}

\newcommand{\sumn}{\sum_{a=1}^n}
\newcommand{\M}{\mathcal{M}}
\newcommand{\qt}{\tilde{q}}
\newcommand{\dpart}[2]{\frac{\partial #1}{\partial #2}}
\newcommand{\thav}[1]{\left< #1 \right>}

\newcommand{\thavn}[1]{\left< #1 \right>_n}

\DeclareMathOperator{\Tra}{Tr}

\newcommand{\beq}{\begin{equation}} 
\newcommand{\eeq}{\end{equation}}
\newcommand{\ba}{\begin{eqnarray}}
\newcommand{\ea}{\end{eqnarray}}

\begin{document}

\title{Low-frequency vibrational spectrum of mean-field disordered systems}

  \author{Eran Bouchbinder}
\affiliation{Chemical and Biological Physics Department, Weizmann Institute of Science, Rehovot 7610001, Israel}
\author{Edan Lerner}
 \affiliation{Institute of Theoretical Physics, University of Amsterdam, Science Park 904, 1098 XH Amsterdam, the Netherlands}
  \author{Corrado Rainone}
\affiliation{Institute of Theoretical Physics, University of Amsterdam, Science Park 904, 1098 XH Amsterdam, the Netherlands}
   \author{Pierfrancesco Urbani}
 \affiliation{Universit\'e Paris-Saclay, CNRS, CEA, Institut de physique th\'eorique, 91191, Gif-sur-Yvette, France}
 \author{Francesco Zamponi}
 \affiliation{Laboratoire de Physique de l'Ecole Normale Sup\'erieure, ENS, Universit\'e PSL, CNRS, Sorbonne Universit\'e, Universit\'e de Paris, F-75005 Paris, France}

 \begin{abstract}
We study a recently introduced and exactly solvable mean-field model for the density of vibrational states $\DD(\omega)$ of a structurally disordered system. 
The model is formulated as a collection of disordered anharmonic oscillators, with random stiffness $\k$ drawn from a distribution $p(\k)$,
subjected to a constant field $h$ and interacting bilinearly with a coupling of strength $J$. 
 We investigate the vibrational
properties of its ground state at zero temperature. When $p(\k)$ is gapped, the emergent $\DD(\omega)$ is also gapped, for small $J$. Upon increasing $J$, 
the gap vanishes on a critical line in the $(h,J)$ phase diagram, whereupon replica symmetry is broken.
At small $h$, the form of this pseudogap is quadratic, $\DD(\omega)\sim\omega^2$, and its modes are delocalized, as expected from previously investigated mean-field spin glass models. However, we determine that for large enough $h$, a quartic pseudogap $\DD(\omega)\sim\omega^4$, populated by localized modes, emerges, the two regimes being separated by a special point on the critical line.
We thus uncover that mean-field disordered systems can generically display both a quadratic-delocalized and a quartic-localized spectrum at the glass transition.
\end{abstract}
 
\maketitle

\paragraph*{Introduction ---}
The vibrational spectrum of structural glasses displays a series of universal features in different frequency ranges, which are responsible 
for important material properties, such as wave attenuation, heat transport, and plasticity~\cite{ruocco2001high,nakayama2002boson,soft_potential_model_1991}. 
Motivated by these observations, several authors have
constructed simple models of the non-phononic vibrational density of states of structurally disordered systems, $\DD(\omega)$~\cite{kuhn_and_Horstmann_prl_1997,Chalker_prb_2003,GPS03,grigera2003phonon,GPS07,eric_boson_peak_emt,FPUZ15,baityjesi2015,SYM16,Gabriele_pre_2018,fyodorov2018hessian,manning_random_matrix_pre_2018,Harukuni_pre_2019,BZ19,Atsushi_soft_matter_dipole_dos,fyodorov2020manifolds,itamar_GPS_prb_2020}.
Mean-field models typically display a quadratic spectrum, $\DD(\omega)\sim \omega^2$~\cite{FPUZ15,SYM16}, of delocalized and featureless modes~\cite{LivanRMT}; this delocalization is inherently different from the one associated with phononic excitations in solids (which are absent in the mean-field limit), and is a manifestation of the marginal stability associated to replica symmetry breaking~\cite{parisiSGandbeyond,SimpleGlasses2020}.

On the contrary, numerical simulations of model glass formers in finite dimension have revealed that non-phononic excitations in those systems are \emph{quasi-localized} in nature, that they emerge from self-organized glassy frustration, and that they follow a seemingly universal \emph{quartic} law ${\cal D}(\omega)\sim\omega^4$~\cite{Schober_Laird_numerics_PRL,modes_prl_2016,MSI17,modes_prl_2018,WNGBSF18,modes_prl_2020}. Given these discrepancies with the mean-field scenario detailed above, and the localized nature of these excitations, 
the naive expectation is that the modes that populate the quartic law would disappear in the mean-field limit, and that mean-field models are therefore unable to tell much about the physics responsible for the $\omega^4$ spectrum of structural glasses~\cite{non_debye_prl_2016,Harukuni_arXiv_2020_mean_field_vdos,atsushi_large_d_packings_pre_2020}.

Nearly two decades ago, Gurevich, Parshin and Schober (GPS) proposed  a three-dimensional lattice model~\cite{GPS03} for this glassy density of states,
formulated in terms of interacting anharmonic oscillators, with a coupling strength that decays with distance as $\sim\!r^{-3}$, 
where $r$ is the distance between the oscillators. GPS showed numerically that ${\cal D}(\omega)\!\sim\!\omega^4$ emerges in that model,
and proposed a phenomenological theory~\cite{GPS03,GPS07}, 
which has later been investigated by other authors~\cite{itamar_GPS_prb_2020}. 
A similar mean-field model was studied by K\"uhn and Horstmann (KH)~\cite{kuhn_and_Horstmann_prl_1997}, who however did not investigate the vibrational spectrum; as we will show below, their model's spectrum follows ${\cal D}(\omega)\!\sim\!\omega^2$.

In this Letter, we study a recently introduced model~\cite{rainone2020solvable}, which corresponds to both the infinite-dimensional, mean-field version of the GPS model, and to a generalization of the KH model. Following Ref.~\cite{rainone2020solvable}, we hereafter refer to it as the \emph{KHGPS model}. The model is formulated as a collection of $N$ interacting anharmonic oscillators, each represented by a generalized coordinate $x_i$ and stiffness $\kappa_i$~\footnote{For our abstract mathematical model, all quantities are assumed to be dimensionless.}; the model's Hamiltonian reads
\beq
 H\equiv \sum_{i < j}J_{ij}x_i x_j + \frac{1}{2}\sum_i \kappa_i x_i^2 + \frac{1}{4!}\sum_i x_i^4 - h\sum_i x_i\,.
 \label{eq:model}
\eeq
Here the interactions $J_{ij}$ are assumed to be Gaussian, i.i.d.~random couplings of variance $J^2/N$ $\forall\ i\!\neq\!j$, and
 $J$ represents the strength of the disordered interactions, taken to be space-independent. The harmonic stiffnesses $\kappa_i$ are characterized by a distribution 
 $p(\kappa)$ which we take as uniform in $[\kappa_m,\kappa_{_{\!M}}]$ with $\kappa_{_{\!M}} \geq \kappa_m \geq 0$, in such a way that all the oscillators have a single minimum at $J=0$. 
 An external constant ``magnetic'' field $h$ is added in order to break the spurious $x_i\to -x_i$ symmetry that has no counterpart in amorphous solids~\cite{BU15, albert2020searching}. 
 The model is related to a soft-spin version of the Sherrington-Kirkpatrick model~\cite{sozi82}.
 
In what follows we describe the exact solution of the KHGPS model using the replica method~\cite{kuhn_and_Horstmann_prl_1997,parisiSGandbeyond}.  We construct the phase diagram
of the model, in the plane of the applied magnetic field $h$ and the coupling strength $J$.
We rigorously show that, for $\k_m>0$, the model's spectrum is gapped at small enough coupling, in the replica symmetric phase where the energy landscape is convex. Upon increasing the coupling strength, a phase transition is encountered, whereupon replica symmetry is broken, the energy landscape becomes rough, and the gap in the spectrum closes.
On this critical line, the spectrum behaves as $\DD(\omega)\sim \omega^2$ at small $h$, a typical mean-field scenario. Conversely, for large $h$, the spectrum behaves as ${\cal D}(\omega)\sim\omega^4$ and its modes are partially localized. The two regimes are separated by a special point on the critical line, whose location is determined.
All in all, our work demonstrates that disordered mean-field models can display a quartic density of states of localized modes in certain regions of their phase diagram, including critical lines whereupon replica symmetry is broken. Related results have been reported in Ref.~\cite{lupo2017critical} for the XY model defined on a random graph, which is however much more difficult to analyze.
This result opens new perspectives for the microscopic understanding of the universal ${\cal D}(\omega)\sim\omega^4$  law in finite-dimensional glassy systems. Furthermore, it shows that replica symmetry
breaking (RSB) phase transitions can present profoundly different characteristics from the marginal stability scenario usually associated to it, even at a mean-field level.

 \paragraph*{Vibrational spectrum --- } 
The Hessian ${\cal M}_{ij}\!\equiv\!\partial^2H/\partial x_i\partial x_j$ corresponding to $H$ takes the form
\begin{equation}
\M_{ij} = J_{ij} + \delta_{ij}\left(\kappa_i + \frac{1}{2}x^2_i\right) \equiv J_{ij}+\delta_{ij}a_i\,,
\label{eq:hessian}
\end{equation}
which is the sum of a member ($J_{ij}$) of the Gaussian Orthogonal ensemble (GOE) of random matrices~\cite{LivanRMT} and of a diagonal matrix $A_{ij}=a_i \d_{ij}$,
with diagonal elements $a_i\equiv\kappa_i + x_i^2/2$.
Assuming that there is no statistical correlation between these two matrices, calculating the spectrum of their sum becomes a standard problem in random matrix theory~\cite{LivanRMT}, which only requires knowledge of the statistics of the diagonal part. Past efforts in calculating typical ground-state spectra of mean-field disordered systems~\cite{FPUZ15} indicate that this assumption is valid, hence we adopt it here and proceed as follows.

Assuming that the statistics $p(a)$ of the diagonal elements is known and has support in $[a_m,a_{_{\!M}}]$ 
 (we will compute it below), one can compute the density of eigenvalues $\rho_{_{\!\M}}(\lambda)$ of $\MM$ by defining the resolvent~\cite{BBP17},
\beq
\mathfrak{g}_{_{\!\M}}(z) \equiv \int \dd \l \ \frac{\rho_{_{\!\M}}(\l)}{z-\l} \,,
\label{eq:def_resolv}
\eeq
which implies
\beq
\rho_{_{\!\M}}(\lambda) = \frac{1}{\pi}\lim_{\eta \to 0^+} \Im [ \mathfrak{g}_{_{\!\M}}(\lambda - i\eta )] \,. 
\label{eq:spec_from_res}
\eeq
The resolvent of $\MM$ is then implicitly expressed in terms of the spectrum of the  
diagonal part, ${\rho_{_{\!A}}(a)= p(a)}$, as~\cite{BBP17}
\beq
\mathfrak{g}_{_{\!\M}}(z) = \int_{a_m}^{a_{_{\!M}}} \dd a \, p(a) \frac{1}{z-a- J^2\mathfrak{g}_{_{\!\M}}(z)} \,.
\label{eq:ris_edge}
\eeq
This equation requires a numerical solution, but the position and shape of the lower edge of the spectrum can be worked out analytically. Let us define  
${g(z)\!\equiv\!\mathfrak{g}_{_{\!\M}}(z)\!-\!z/J^2}$; one can then recast Eq.~\eqref{eq:ris_edge} as 
\beq
z = - J^2 \int_{a_m}^{a_{_{\!M}}} \dd a \, p(a) \left[ g + \frac{1}{a+ J^2 g}  \right] \equiv \FF(g) \,.
\eeq

We next argue as follows: for $\lambda$ outside of the support of the spectrum $\rho_{_{\!\M}}(\lambda)$ and in the limit $\eta\!\to\!0$, Eq.~\eqref{eq:spec_from_res} implies that $g(z)$ cannot have an imaginary part. We therefore expect a band of values of the function $\FF(g)$ to be forbidden for real $g$, and to correspond to the support of the spectrum. Let us then consider which values $\FF(g)$ can attain for real $g$. 
This function is obviously not defined to the left of $-a_{_{\!M}}/J^2$ or to the right of $-a_m/J^2$, 
  and intuitively, we expect the branch for $g\!>\!-a_m/J^2$ to be the one controlling the \emph{lower} edge; therefore, this branch needs to be bounded from above. There are then only two possibilities:
(i) The function $\FF(g)$ has a maximum $g_m$ for $g\!>\!-a_m/J^2$, meaning that $\FF'(g_m)\!=\!0$. The corresponding value of $\FF$, $\l_m\!=\!\FF(g_m)$, is then the lower edge of the spectrum. In this case, the support of the diagonal elements has no influence on the lower edge, and the GOE part of $\MM$ dominates: close to the edge the eigenvectors are delocalized and $\rho_{_{\!\M}}(\lambda)\!\propto\!(\lambda\!-\!\lambda_m)^{1/2}$~\cite{LS16}. We dub this a GOE-like spectrum.
(ii) The function $\FF(g)$ has no maximum for $g\!>\!-a_m/J^2$. In this case, the value of $g$ that corresponds to the edge must be $g_m\!=\!-a_m/J^2$, and the lower edge itself is 
\beq
\l_m = \FF\left(-\frac{a_m}{J^2}\right) = a_m - J^2 \int_{a_m}^{a_{_{\!M}}} \dd a \, p(a)  \frac{1}{a- a_m}  \ .
\label{eq:edge}
\eeq
The edge is then determined by the support of $p(a)$, and dominated by the diagonal part of $\MM$,
and the eigenvectors near the edge are partially localized~\cite{LS16}. We dub this a DIAG-like spectrum.
Furthermore, if $p(a) \sim (a-a_m)^\nu$ near its lower edge,
one can show by an expansion near $g_m$ that $\r_{_{\!\M}}(\lambda)\sim (\l-\l_m)^\nu$.

The value of coupling that separates the two regimes is such that $\FF'(g=g_m)=0$, which gives the self-consistent equation:
\beq\label{eq:Jc}
\L = 1 - J^2\int_{a_m}^{a_{_{\!M}}} \dd a \, p(a)  \frac{1}{(a -a_m)^2}  = 0 \ ,
\eeq
such that $\L>0$ corresponds to a DIAG-like and $\L<0$ corresponds to a GOE-like spectrum.
Note that because $p(a)$ and its edges depend on $J$, this equation defines the critical value of coupling at which the spectrum changes 
shape only implicitly.

\paragraph*{Replica method ---}
We next aim at determining the statistics $p(a)$ of the diagonal elements $a_i$ appearing in Eq.~\eqref{eq:hessian}, which depend on the oscillator positions $x_i$ in the ground state of the model, and can be determined by solving its thermodynamics in the zero-temperature (${T\!\to\!0}$) limit. We do so by employing the replica method~\cite{parisiSGandbeyond}; we assume that the ground state is unique, which corresponds to a replica-symmetric (RS) ansatz. This picture is expected to be justified as long as the coupling strength $J$ is below a critical threshold $J_c(h)$, which we self-consistently determine below.

We briefly delineate the steps in obtaining the RS solution of the model, leaving the details to Appendix~\ref{app:B}. 
The mean-field nature of the model allows one to write its solution as
 a problem of decoupled oscillators in an effective, self-consistent random external potential, which in the $T\!\to\!0$ limit takes the form
\beq
v_{\rm eff}(x) \equiv \frac{x^4}{4!} + \frac{m}{2}x^2 - (f+h) x \ , \quad m =\kappa-J^2 \c \ ,
\label{eq:veff_zero_T}
\eeq
where $f$ is a Gaussian random force of zero mean and variance $J^2\tilde{q}$,
and the new parameters $\chi$ and $\qt$ emerge from the correlations between different replicas generated by the disorder,
and have to be determined self-consistently~\cite{parisiSGandbeyond}. 

Depending on the value of the coefficients, the effective potential can be either an asymmetric single well (SW) or double well (DW), with two minima separated by an energy barrier.
In particular, if the effective stiffness $m$ is negative, there is always some value of the field $f$ for which the potential is a DW. We thus conclude that if
${m_m = \kappa_m - J^2\chi < 0}$, DWs appear with finite probability, and we show in Appendix~\ref{app:Tzero} that in this case the RS solution is always unstable towards RSB. 
Consequently, we now restrict ourselves to the case $m_m \geq 0$, which is realized at small enough $J$ if $\kappa_m>0$, and we discuss the RS phase of the model.

Under the assumption $m_m\geq 0$, we show in the Appendix that
the parameters $\tilde q$ and $\chi$ are self-consistently determined through the equations
\beq\label{eq:sp_zero_T}
\chi = \left\langle \frac{1}{ v_{\rm eff}''(x^*(f,m))} \right\rangle_{m,f} \ , \quad
\qt = \langle \left(x^*(f,m)\right)^2  \rangle_{m,f} \ ,
\eeq
where $x^*(f,m)$ denotes the point of absolute minimum of the effective potential, and the average is taken over the
random effective stiffnesses $m\,$$\,\sim\,$$\,U(m_m$$=$$\kappa_m$$-$$J^2\c,\,$$m_{_{\!M}}$$=$$\kappa_{_{\!M}}$$-$$J^2\c)$ and 
random fields ${f \sim \NN(0,J^2 \tilde q)}$.
The self-consistency of this picture is tested by verifying the positivity of the \emph{replicon} eigenvalue $\l_R$ of the Hessian matrix of the replica action~\cite{parisiSGandbeyond,SimpleGlasses2020}. The definition of the replica action and the computation of the replicon can be found in Appendix~\ref{app:replicon}. The final result reads
\beq
\l_R = 1 - J^2\left\langle \frac{1}{v_{\rm eff}''(x^*(f,m))^2} \right \rangle_{m,f} \ .
\eeq
Recalling that 
${a = \kappa + x^2/2 = v_{\rm eff}''(x) + J^2 \c}$, cf. Eq.~\eqref{eq:hessian}, where $x$ has to be evaluated in $x^*(f,m)$, we
can express $\c$ and $\l_R$ as
\beq\begin{split}
\chi &= \int_{a_m}^{a_{_{\!M}}} \dd a \, p(a)  \frac{1}{a -J^2 \c} \ , \\
\l_R &= 1 - J^2 \int_{a_m}^{a_{_{\!M}}} \dd a \, p(a)  \frac{1}{(a -J^2 \c)^2} \ ,
\label{eq:replicon_zero_T}
\end{split}\eeq
where $\l_R$ differs from $\L$ in Eq.~\eqref{eq:Jc} by the replacement of $a_m$ by $J^2 \c$ in the denominator of the integrand.
Note that ${a = m + J^2 \c + x^2/2}$, hence under the assumption that $m_m\geq 0$, we have
${a_m \geq J^2\c}$, which imples $\l_R \geq \L$.

\begin{figure}[t]
\centering
\includegraphics[width=0.5\textwidth]{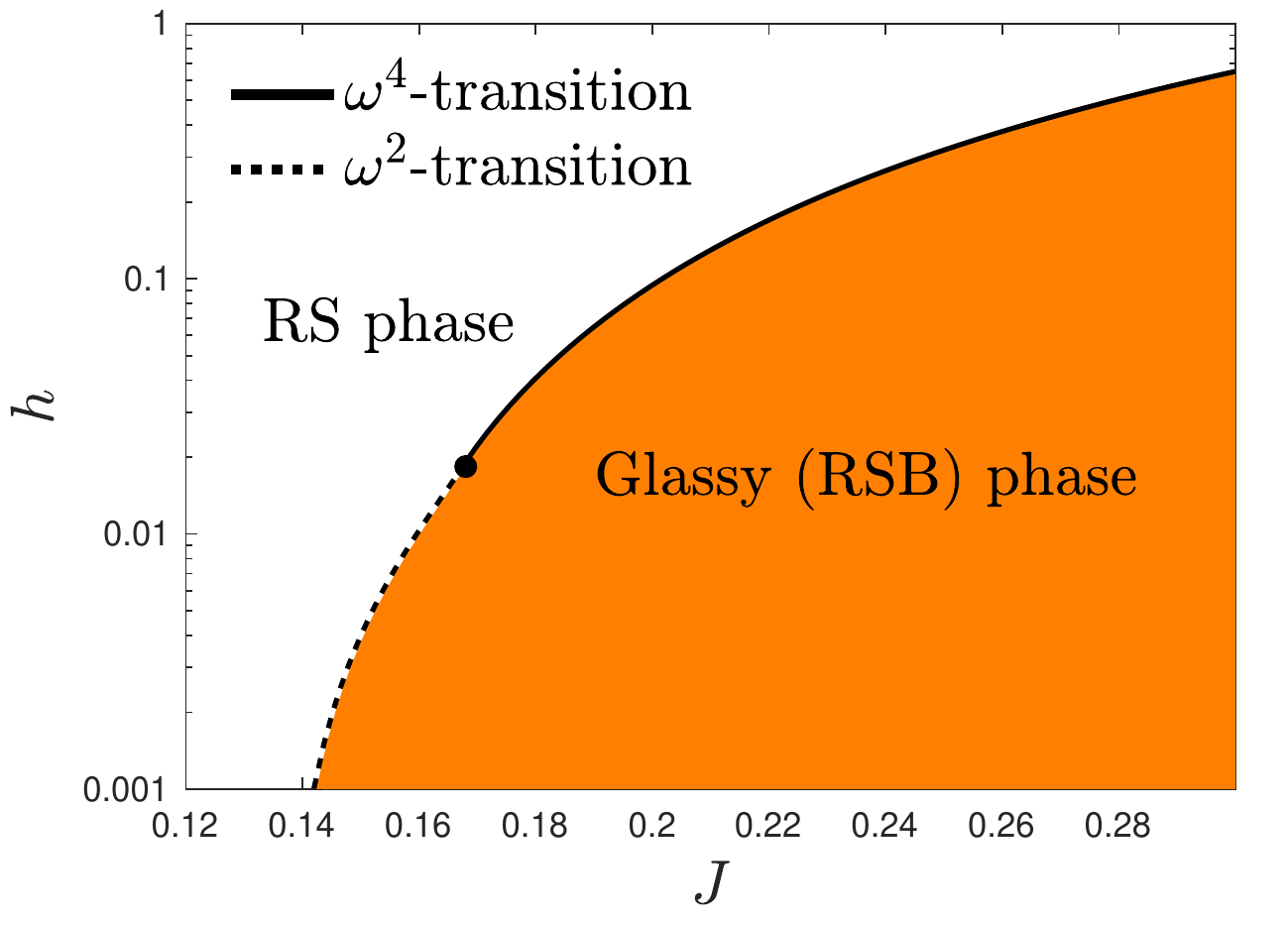}
\caption{Phase diagram of the model for $\k_m=0.1$ and $\k_{_{\!M}}=1$ in the $(h,J)$ plane. The line $J_c(h)$ separates the convex-landscape RS phase from
the rough-landscape RSB phase.
Along the dotted line, $\mathcal{D}(\omega)\propto\omega^2$ at the transition, whilst $\mathcal{D}(\omega)\propto\omega^4$ on the solid line.}
\label{fig:PD}
\vskip-10pt
\end{figure}

 \paragraph*{Phase diagram ---}
We are now in a position to determine the phase diagram of the model in the $(h,J)$ plane.
We fix here $\kappa_m=0.1$ and $\kappa_{_{\!M}}=1$, but the qualitative picture is independent of this choice as long as $\k_m>0$.
At $J=0$, the curvatures $v_{\rm eff}''(x^*)$ are finite and consequently $\c$ is finite. Hence, at $J=0$ we have $m=\kappa$ and consequently
for small enough $J$, the condition $m_m > 0$ is satisfied. Because $m_m>0$, we have $a_m > J^2 \c$ and the integral
that appears in Eq.~\eqref{eq:replicon_zero_T} is finite, leading to $\l_R \approx 1$ for $J\approx 0$. We thus conclude that the RS phase is
stable at small $J$. While it is easy to show that the spectrum is always gapped in this phase, 
the sign of $\L$, and thus the shape of the spectrum near its edge, 
depends on the behavior of $p(a)$ near $a_m$ and on the values of $a_m$ and $a_{_{\!M}}$.
For example, the KH model studied in Ref.~\cite{kuhn_and_Horstmann_prl_1997} has $\kappa_m=\kappa_{_{\!M}}=1$ and in that case $a_m=a_{_{\!M}}$ leading
formally to $\L = -\io$, in such a way that the spectrum is always GOE-like. With our choice of $p(\kappa)$, instead, the integral in Eq.~\eqref{eq:Jc}
is finite and the spectrum is always DIAG-like at low enough $J$.

The RS phase can then become unstable in two ways:
(i) The replicon can vanish, while $m_m$ remains strictly positive. In this case, at the transition point
we have $\L \leq \l_R =0$, hence the spectrum is GOE-like. 
For a GOE-like spectrum to be
gapless, the two equations $\l_m$$=$$\FF(g_m)$$=$$0$ and $\FF'(g_m)$$=$$0$ must hold, which is equivalent to Eqs.~\eqref{eq:replicon_zero_T}
with $\l_R$$=$$0$ and $g_m$$=$$-\c$. Hence, the spectrum is gapless at the critical point and $\r_{_{\!\M}}(\l) \sim \l^{1/2}$, which
is equivalent to $\DD(\omega)\sim \omega^2$. Just above the critical point, the replicon becomes negative. This is a standard
RSB transition, observed in several spin glass models.
(ii) The lower bound of the effective stiffness can vanish, $m_m=0$, while the replicon is still positive, $\l_R>0$.
When $m_m < 0$, there is a finite probability of having DWs in the ensemble of effective potentials, and we show in Appendix~\ref{app:Tzero} that this
formally implies $\l_R = -\io$. Hence, the replicon jumps discontinuously to minus infinity beyond this transition. At the transition point,
$m_m=0$ implies (see Appendix~\ref{sec:DW} for details) that $a_m=J^2\c$, which implies that $\L = \l_R >0$ and the spectrum is DIAG-like.
Furthermore, close to its lower edge,
\beq
p(a) \sim (a - J^2 \c)^{3/2} \qquad \Longrightarrow \qquad p(\tilde a )\sim \tilde a^{3/2} \ ,
\eeq
where $\tilde a = v_{\rm eff}''(x^*) = a - J^2 \c$ is the curvature of the effective potential at its minimum.
Note that Eq.~\eqref{eq:sp_zero_T} then gives $\chi = \langle 1/\tilde a\rangle$ and from 
Eq.~\eqref{eq:edge} it follows that $\l_m = J^2 [\chi - \langle 1/\tilde a \rangle]=0$. 
We conclude that the spectrum is gapless and DIAG-like, i.e. $\r_{_{\!\M}}(\l)\sim\l^{3/2}$, or equivalently
$\DD(\omega)\sim \omega^4$.

\begin{figure}[t]
 \centering
 \includegraphics[width=0.45\textwidth]{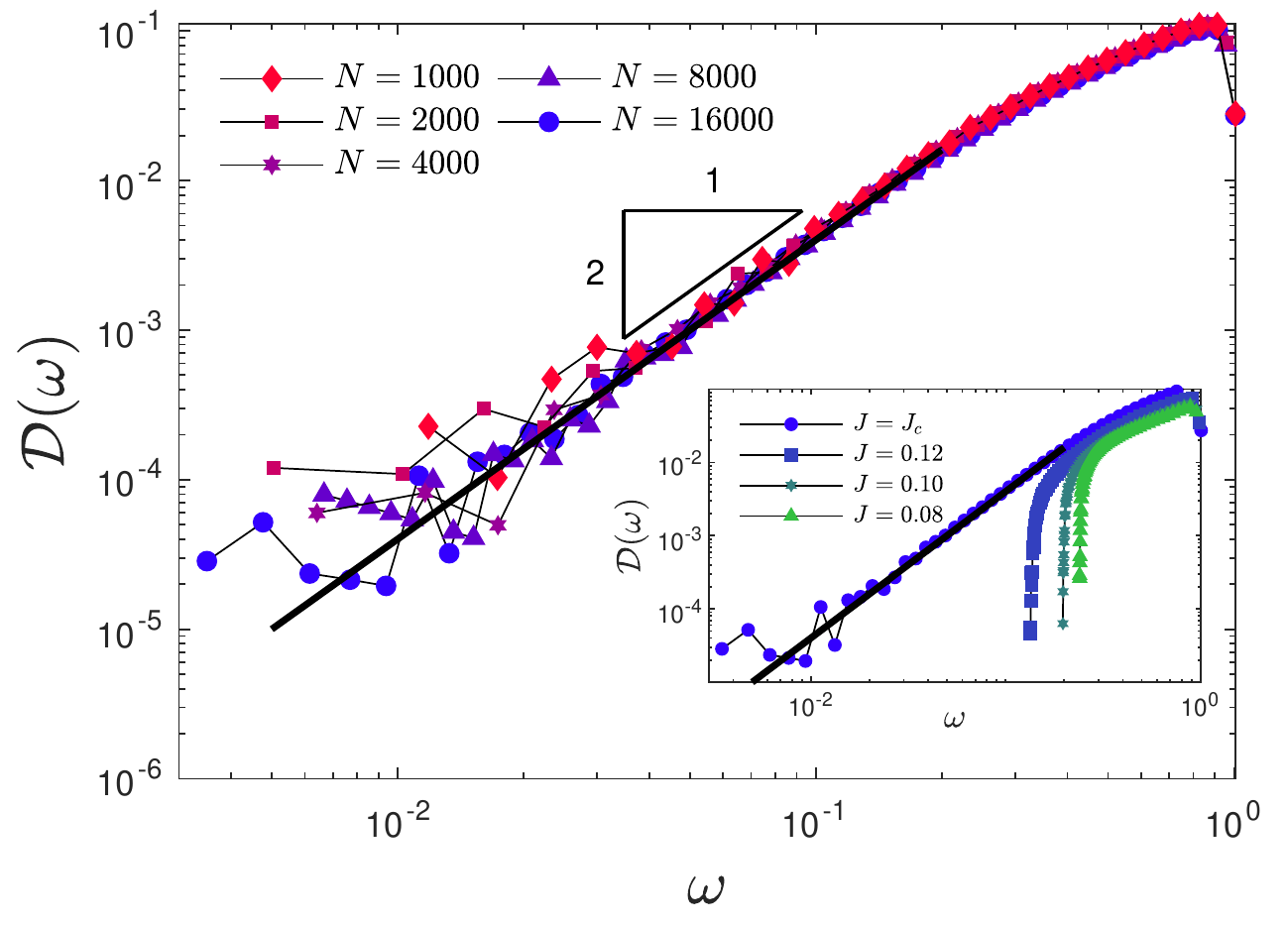}
 \includegraphics[width=0.45\textwidth]{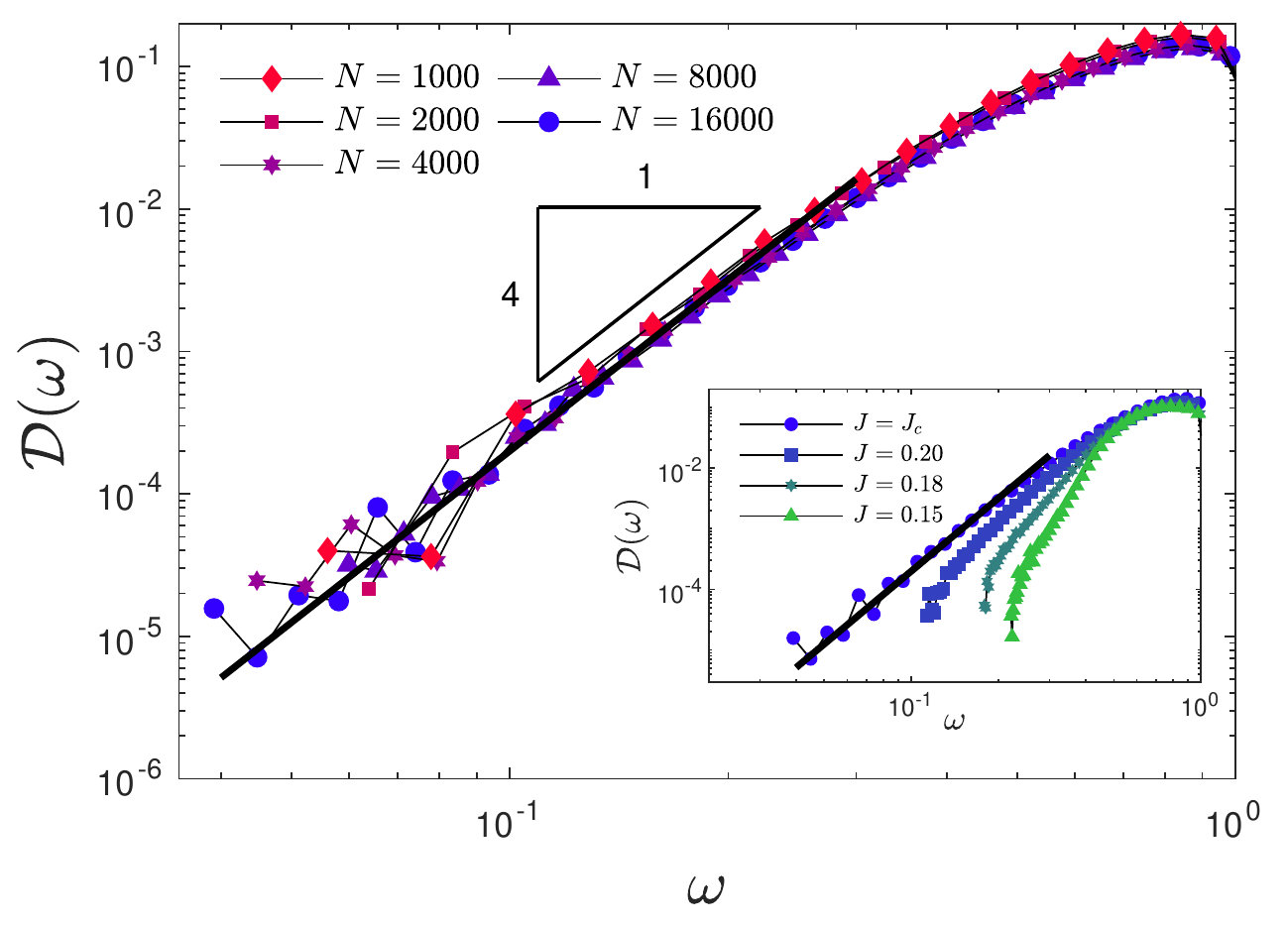}
 \vskip-10pt
 \caption{Numerical results for the vibrational spectrum of the model, at two selected values of $h$ 
 (top: $h=0$, ${J_c(h)=0.137138}$ and bottom: $h=0.157220$, $J_c(h)=0.218$)
 corresponding to the two possible shapes of the spectrum. 
 At the transition, we show multiple system sizes to confirm the gaplessness of the spectrum (main panels).
 The gap in the spectrum closes as the critical line is approached from the RS phase by increasing the coupling (insets). 
 }
 \label{fig:spectrum}
 \vskip-10pt
 \end{figure}

The phase diagram obtained by solving numerically Eqs.~\eqref{eq:sp_zero_T} is reported in Fig.~\ref{fig:PD}. We observe
that the glass transition line $J_c(h)$ falls into case (i) for small $h$, and into case (ii) for large $h$. The two lines are separated
by a special point at which $m_m=0$ and $\l_R=0$ simultaneously. We also verify these predictions numerically, by directly calculating the spectrum $\mathcal{D}(\omega)$ of the Hessian in the minima of the Hamiltonian in Eq.~\eqref{eq:model}, obtained by means of a gradient descent algorithm~\footnote{We use the BFGS algorithm implemented in the GSL library~\cite{GSL_library}.}. These numerical results, which confirm our theoretical predictions, are reported in Fig.~\ref{fig:spectrum}.
We note that when $\k_m$ is reduced and approaches zero, the line $J_c(h)$ moves towards the left, i.e. towards smaller values of $J$, and the $\omega^4$ region increases; when
$\k_m=0$, the model is in the RSB phase at all $J$. This regime was studied numerically and through a scaling theory in Ref.~\cite{rainone2020solvable}.

\paragraph*{Discussion ---}
We studied a mean-field model of interacting disordered anharmonic oscillators~\cite{rainone2020solvable} having, in absence of coupling, a gapped spectrum.
We showed that at small coupling the spectrum remains gapped~\cite{ji2020thermal},
and that at the glass transition point it can display either 
the universal ${{\cal D}(\omega)\!\sim\!\omega^4}$ localized spectra observed in finite-dimensional computer glass models~\cite{baityjesi2015,modes_prl_2016,modes_prl_2018,modes_prl_2020,WNGBSF18,MSI17} and in the random graph XY model~\cite{lupo2017critical}, 
or the standard ${{\cal D}(\omega)\!\sim\!\omega^2}$ observed in most
mean-field spin glass models and jammed sphere packings~\cite{SYM16,non_debye_prl_2016,eric_boson_peak_emt,FPUZ15}. The immediate implication of our results is that systems at a RSB transition, and possibly even deep within the RSB phase, can in fact exhibit localized excitations, even at the mean-field level. The class of models to which the KHGPS model studied here belongs is expected to be rather broad --- according to existing evidence~\cite{GPS03,itamar_GPS_prb_2020,Chalker_prb_2003} --- and largely robust to changes in these models' input.  

We note that an effective potential in the form of a quartic polynomial, Eq.~(\ref{eq:veff_zero_T}), naturally emerges from our theory. This effective potential, which resembles the Soft Potential Model framework~\cite{soft_potential_model_1991,Schober_prb_1992,GC03} that also predicts a $\omega^4$ nonphononic spectrum under some nontrivial assumptions (spelled out, e.g., in~\cite{Chalker_prb_2003}). In light of our results, the Soft Potential Model can be viewed as an effective description of the collective, many-body statistical-mechanics of the KHGPS model. 

Moreover, we note that the zero temperature limit of the spin glass susceptibility behaves very differently on the two parts of the critical line, being divergent when the spectrum at the transition is $\omega^2$, and finite when the spectrum is $\omega^4$~\footnote{The spin glass susceptibility is defined by $\chi_{SG}=\sum_{ij}\overline{\left(\langle x_i x_j \rangle-\langle x_i\rangle \langle x_j \rangle\right)^2}/N$, and in the zero temperature limit it becomes $\chi_{SG} =\frac{1}{N}\textrm{Tr}\MM^{-2} $.}.
 Finally, we stress that our results apply upon approaching the transition at strictly zero temperature, 
 and therefore it is important to investigate the model's behavior at finite temperature.

In this work, we limited ourselves to the investigation of the RS phase of the model with $\k_m>0$, up to the critical line whereupon replica symmetry is broken and a glassy phase appears. A natural direction for future research is to investigate the vibrational spectrum deep in the glass phase. One might expect, by continuity arguments, that the quartic spectrum extends into
 the glass phase, hence being valid in a finite region of the phase diagram. This point of view seems to be supported by the numerical results of Ref.~\cite{rainone2020solvable},
but whether this intuition is correct can only be confirmed by an investigation of the RSB equations of the model.
The gradient descent dynamics might also display interesting features in the glass phase~\cite{cuku93,folena2020rethinking} 
and, if minima reached by quenching dynamics retain the properties of the model at the transition, one could expect different (or even the absence of) aging dynamics.

\emph{Acknowledgements}.---We benefited from discussions with Giulio Biroli, Jean-Philippe Bouchaud, Gustavo D\"uring, Eric De Giuli, and Guilhem Semerjian. This project has received funding from the European Research Council (ERC) under the European Union's Horizon 2020 research and innovation programme (grant agreement n° 723955 - GlassUniversality) and by a grant from the Simons Foundation (\#454955, Francesco Zamponi). P.~U.~acknowledges support by "Investissements d'Avenir" LabEx-PALM  (ANR-10-LABX-0039-PALM). E.~B.~acknowledges support from the Minerva Foundation with funding from the Federal German Ministry for Education and Research, the Ben May Center for Chemical Theory and Computation, and the Harold Perlman Family. E.~L.~acknowledges support from the NWO (Vidi grant no.~680-47-554/3259). 

 \vskip-20pt

\bibliography{refs.bib}

\clearpage

\begin{widetext}

\appendix

\section{The spectrum}

The Hessian matrix, evaluated in a minimum $x^*_i$ of the Hamiltonian,
 is the sum of a GOE matrix $J_{ij}$ and of a diagonal matrix $A_{ij} =a_i \d_{ij}$ with diagonal elements
$a_i =  \kappa_i + \frac{1}{2}(x^*_i)^2$.
The random variable $a_i$ is distributed according to $p(a)$ in the interval $[a_m, a_{_{\!M}}]$. 
In the following, we assume for simplicity that $a_i$ and $J_{ij}$ are uncorrelated; 
it can be proven both analytically and numerically that this assumption is correct~\cite{FPUZ15}.

\subsection{Resolvent equation}

We want to calculate the density of eigenvalues $\rho_{_{\!\M}}(\lambda)$ of the matrix $\M$. This can be defined in terms of the resolvent (or rather, the trace of the resolvent in the thermodynamic limit)
\beq
\mathfrak{g}_{_{\!\M}}(z) \equiv \lim_{N\to\infty}\frac{1}{N}\Tra (z\mathbf{1} - \M)^{-1} = \int \dd \l\ \frac{\rho_{_{\!\M}}(\l)}{z-\l} \ ,
\eeq
which implies
\beq
\rho_{_{\!\M}}(\lambda) = \frac{1}{\pi}\lim_{\eta \to 0^+} \Im \mathfrak{g}_{_{\!\M}}(\lambda - i\eta ) \ . 
\eeq
The resolvent of $\M$ can be obtained in terms of the resolvent of the diagonal matrix $A$ via the fixed-point equation
\beq
\mathfrak{g}_{_{\!\M}}(z) = \mathfrak{g}_{A}(z- J^2\mathfrak{g}_{_{\!\M}}(z)),
\eeq
where we used the fact that $J_{ij}$ is a GOE matrix~\cite{BBP17}.
The resolvent of a diagonal matrix is trivial because $\r_{_{\!A}}(\l) = p(a)$, hence the fixed-point equation is explicitly written as
\beq\label{eq:ris_edge_SI}
\mathfrak{g}_{_{\!A}}(z) = \int \dd a \, p(a) \frac{1}{z-a} \qquad
\Rightarrow
\qquad
\mathfrak{g}_{_{\!\M}}(z) = \int \dd a \, p(a) \frac{1}{z-a- J^2\mathfrak{g}_{_{\!\M}}(z)} \ .
\eeq

\subsection{Location of the spectrum edge}

To investigate the low-frequency tail of the spectrum we start from Eq.~\eqref{eq:ris_edge_SI}, and
we define $g(z) = \mathfrak{g}_{_{\!\M}}(z) - z/J^2$ so we get
\beq
z =- J^2 \int \dd a \, p(a) \left[ g + \frac{1}{a+ J^2 g}  \right] \equiv \FF(g) \ .
\eeq
Outside the support of the spectrum, for $z=\l - i \h$ with $\h\to 0$, $g(z)$ needs to be real. We expect that there is a band of values of $\FF(g)$ that
are forbidden for real $g$, which correspond to the support of the spectrum.
So, we study the function $\FF(g)$ for real $g$.
We have
\beq\label{eq:Fg_der}
\FF'(g) = -J^2 \int \dd a \, p(a) \left[ 1 - \frac{J^2}{(a + J^2 g)^2}  \right]  \ , \qquad
\FF''(g) = -J^2 \int \dd a \, p(a) \left[ 2 \frac{J^4}{(a+ J^2 g)^3}  \right]  \ .
\eeq
Note that if $p(a)$ has support in $[a_m,a_{_{\!M}}]$, then $\FF(g)$ is only defined for $g\notin [-a_{_{\!M}}/J^2, -a_m/J^2]$ on the real axis.

There are two possibilities for the spectrum:
\begin{itemize}

\item[GOE-like--]
Suppose that the function $\FF(g)$ has a minimum for $g_{_{\!M}} < -a_{_{\!M}}/J^2$ and a maximum for $g_m > -a_m/J^2$. 
In this case, if $g_m,\ g_{_{\!M}}$ are the solutions of $\FF'(g)=0$, then $\l_m=\FF(g_m)$ and $\l_{_{\!M}}=\FF(g_{_{\!M}})$ are the edges of the spectrum.
In the vicinity of the edges we can expand, e.g. for $z = \l_{_{\!M}} - \e$ and $g = g_{_{\!M}} + \d g$, and we get
\beq
\l_{_{\!M}} - \e \sim \FF(g_{_{\!M}} + \d g) \sim \FF(g_{_{\!M}}) + \frac12 \FF''(g_{_{\!M}}) \d g^2 + \cdots
\quad \Rightarrow \quad \e = - \frac12 \FF''(g_{_{\!M}}) \d g^2 + \cdots 
\quad \Rightarrow \quad 
\d g =\sqrt{-2\frac{\l_{_{\!M}} - z}{\FF''(g_{_{\!M}})}}  \ .
\eeq
Clearly if $\FF''(g_{_{\!M}}) \neq 0$ we get $\r(\l) \sim \sqrt{\l_{_{\!M}} - \l}$ in the vicinity of $\l_{_{\!M}}$. The same happens for the lower edge.

\item[DIAG-like--]
It can happen however that $\FF(g)$ has no maximum for any $g > -a_m/J^2$. In this case, the value of $g$ that corresponds to the 
edge is $g_m = -a_m/J^2$, and the location of the edge is
\beq\label{eq:edge2}
\l_m = \FF(-a_m/J^2) = a_m - J^2 \int_{a_m}^{a_{_{\!M}}} \dd a \, p(a) \left[  \frac{1}{a- a_m}  \right] \ .
\eeq

\item[Critical $J$--]
The value of coupling that separates the GOE and DIAG regimes is such that $\FF'(g=g_m)=0$, which gives the self-consistent equation:
\beq
\L = 1 - J^2\int_{a_m}^{a_{_{\!M}}} \dd a \, p(a)  \frac{1}{(a -a_m)^2}  = 0 \ .
\eeq
Note that because $p(a)$ and its edges depend on $J$, this equation defines the critical value at which the spectrum changes 
shape only implicitly. The case $\L>0$ corresponds to $\FF'(g=g_m)<0$, hence to a DIAG-like spectrum, while the case
$\L<0$ corresponds to $\FF'(g=g_m)>0$, hence to a GOE-like spectrum.
\end{itemize}

\subsection{Shape of the edge and prefactors}

We now focus on the DIAG-like spectrum and we study in more details the behavior close to the edge.
The analysis depends on the details of $p(a)$, so we will assume the power-law form
\beq
p(a) \sim \AA_d (a-a_m)^{3/2}
\eeq
close to the lower edge. The analysis is performed similarly for other values of the exponent $\nu \neq 3/2$.

We know that the maximum of $\FF(g)$ is in $g_m = - a_m/J$ and we want to expand $\FF(g)$ around it. 
From Eq.~\eqref{eq:Fg_der} we observe that $\FF'(g_m) = -J^2 \L$ is finite, while $\FF''(g_m)$ is divergent, which suggests a non-analytic behavior for $\FF(g)$
with exponent $3/2$ around $g_m$, as we now show.
We define $\d g = g -g_m$ and
\beq
\d \FF(\d g) = \FF(g_m + \d g) - \FF(g_m) - \FF'(g_m) \d g = 
- J^6 \d g^2 \int \dd a \, p(a) \frac{1}{(a - a_m)^2 (a- a_m + J^2 \d g)} \ .
\eeq
For small $\d g$ we have, defining $q(a) = p(a)/(a-a_m)^{3/2} \to \AA_d$ for $a\to a_m$,
and changing variable to $x = (a-a_m)/\d g$,
\beq
\BB =\lim_{\d g\to 0} \sqrt{\d g} \int \dd a \frac{q(a)}{\sqrt{a - a_m} (a- a_m + J^2 \d g)}=\lim_{\d g\to 0}\int \dd x  \frac{q(a_m+x\d g)}{\sqrt{x} (x+J^2)} = \AA_d \frac{\pi}J \ .
\eeq
Collecting all together these results, we have for small $\d g$:
\beq
\d z= z - \l_m =  \FF(g) - \FF(g_m) \sim  - J^2 \L \d g - J^5 \pi \AA_d \d g^{3/2} + \cdots \ .
\eeq
Inverting this relation we obtain
\beq
\d g(z) = -\frac1{J^2 \L} \d z -\frac{ \pi \AA_d}{ \L^{5/2}} ( -\d z )^{3/2} + \cdots \ .
\eeq
If we choose $\d z = \d \l$ to be real and positive, we get
\beq
\Im \mathfrak{g}_{_{\!\M}}(z) = \Im g(z) = \frac{ \pi \AA_d}{ \L^{5/2}} \d\l^{3/2}  \ , \qquad
\Rightarrow \qquad
\AA_g = \frac{ \AA_d}{ \L^{5/2}} \ .
\eeq
Similar results are obtained for other values of $\n$.

\subsection{Summary}

So far, we have obtained the following results, for a yet unknown $p(a)$:
\begin{itemize}
\item There exist a critical value of coupling (or of other parameters) defined by the condition ${\L = 1 - J^2 \mathbb{E}[\frac{1}{(a-a_m)^2}]=0}$, which separates a DIAG-like spectrum from a GOE-like spectrum.
\item When $\L<0$ the spectrum is GOE-like, the lower edge is given by the solution $g_m$ of $\FF'(g)=0$ and $\l_m =\FF(g_m)$. The spectrum is $\r(\l) \sim \sqrt{\l - \l_m}$ close
to the edge.
\item When $\L>0$ the spectrum is DIAG-like, i.e. it is dominated by the distribution of diagonal elements $p(a)$. Assuming $p(a)\sim \AA_d (a-a_m)^{3/2}$, we find that the lower edge is
$\l_m = a_m - J^2 \mathbb{E}  \left[  \frac{1}{a- a_m}  \right]$ and $\r(\l) \sim \AA_g (\l - \l_m)^{3/2}$ with $\AA_g =  \AA_d/\L^{5/2}$.
\end{itemize}
We now need to obtain information on $p(a)$, i.e. on the statistics of $x^*_i$ in the minima of the Hamiltonian. We do so by solving the thermodynamics of the model in the $T\to 0$ limit.

\section{Replica-symmetric solution of the model}
\label{app:B}

\subsection{The partition function and the free energy}

The replicated partition function at finite temperature $T = 1/\b$ (the Boltzmann constant is set to $k_B=1$), after having averaged over the disorder in the couplings $J_{ij}$ and stiffnesses $\kappa_i$ (whose distribution $p(\k)$ we leave unspecified for now), and introduced the overlap matrix $Q_{ab}$, is~\cite{parisiSGandbeyond}
\beq
\overline{Z^n} = \int  \dd Q_{ab} e^{-\frac{(\beta J)^2}{4}N\sum_{ab}^n Q^2_{ab}} \left[\int \dd p(\kappa)\int \dd^n x \exp\left[-\frac{\beta\kappa}{2}\sumn x_a^2 + \beta h \sumn x_a  - \frac{\beta }{4!}\sumn x_a^4 + \frac{(\beta J)^2}{2}\sum_{ab}^n x_aQ_{ab}x_b\right]\right]^N \ .
\eeq
We now assume a RS form for the $Q_{ab}$ matrix,
$Q^{\rm RS}_{ab} \equiv (\qt - q)\d_{ab} + q$,
which gives for $\overline{Z^n}$
\beq
\begin{split}
\overline{Z^n}& =\int \dd Q^{\rm RS}_{ab} e^{-\frac{(\beta J)^2}{4}Nn [\qt^2 + (n-1)q^2]}\\
& \times\left[\int\dd p(\kappa) \int \dd^n x \exp\left[-\frac{\beta\kappa}{2}\sumn x_a^2 + \beta h \sumn x_a - \frac{\beta }{4!}\sumn x_a^4 + \frac{(\beta J)^2}{2}(\qt-q)\sumn x_a^2 + \frac{(\beta J)^2}{2}q\left(\sumn x_a \right)^2 \right]\right]^N.
\end{split}
\eeq
We rewrite the last term using an Hubbard-Stratonovich transformation
\beq
\exp\left[\frac{(\beta J)^2}{2}q\left(\sumn x^a\right)^2\right]= \int \dd z \frac{1}{\sqrt{2\pi}} \exp\left[-\frac{z^2}{2} + z(\beta J)\sqrt{q}\sumn x^a\right]\equiv \int \DD z \exp\left[z(\beta J)\sqrt{q}\sumn x^a\right],
\eeq
where $z\sim \NN(0,1)$ is a random variable distributed according to a standard normal distribution.
This relation allows us to write
\beq
\overline{Z^n} = \int \dd q \dd \tilde q e^{-\frac{(\beta J)^2}{4}Nn [\qt^2 + (n-1)q^2]}\left[\int \dd p(\kappa)\int \DD z \left(\int \dd x\ e^{-\beta v_{\rm eff}(x)}\right)^n\right]^N,
\eeq
with the definition of the {\it effective potential}:
\beq
v_{\rm eff}(x) \equiv \frac{\kappa}{2}x^2 + \frac{1}{4!}x^4 -\frac{\beta J^2}{2} (\qt-q)x^2 - (J \sqrt{q} z + h)x \ .
\label{eq:veff_app}
\eeq
Note that the random force $f = J \sqrt{q} z$ is Gaussian distributed
with zero mean and variance $J^2 q$.
The replicated partition function can finally be written as
\beq
\overline{Z^n} =\int \dd q \dd \qt e^{-NG(\qt,q)} \ ,
\eeq
with the replica-symmetric action $G$ defined as
\beq
G(\qt,q) = \frac{(\beta J)^2}{4}n[\qt^2 + (n-1)q^2] - \log\left[\int\dd p(\kappa)\int\DD z \left(\int \dd x\ e^{-\beta v_{\rm eff}(x)}\right)^n\right].
\eeq
Assuming that $q$ and $\qt$ have been already selected using the saddle point method, we can then write the replica-symmetric free energy using the replica trick~\cite{parisiSGandbeyond}
\beq
\overline{\log Z} = \lim_{n\to 0}\frac{1}{n} \overline{Z^n}, 
\eeq
which, once the $n\to0$ limit is taken, gives
\beq
f_{\rm RS}(\qt,q) = \frac{(\beta J)^2}{4} (\qt^2-q^2) - \int \dd p(\kappa)\int \DD z \log\left(\int \dd x\ e^{-\beta v_{\rm eff}(x)}\right).
\label{eq:frs}
\eeq

\subsection{Saddle-point equations}
The saddle point equations for $\qt$ and $q$ can be found by differentiating $f_{\rm RS}$, Eq.~\eqref{eq:frs}, with respect to $\qt$ and $q$. The term to the left is trivial, whilst the second requires one to keep in mind the definition Eq.~\eqref{eq:veff_app} of the effective potential $v_{\rm eff}(x)$ and its dependence on $q$ and $\qt$. One gets
\ba
\dpart{f}{\qt} &=& 0 \Longrightarrow \qt = \int \dd p(\kappa) \int \DD z \left<x^2\right>\ ,\label{eq:RSqt}\\
\dpart{f}{q} &=&  0 \Longrightarrow q = \int  \dd p(\kappa) \int \DD z \left<x^2 - \frac{zx}{\sqrt{q}\beta J}\right> \ ,\label{eq:RSq}
\ea
where the bracket $\left<\bullet\right>$ denote a Gibbs average over the effective potential $v_{\rm eff}(x)$,
\beq
\left<\mathcal{O}(x)\right> \equiv \frac{\int \dd x\ \mathcal{O}(x)e^{-\beta v_{\rm eff}(x)}}{\int \dd x\ e^{-\beta v_{\rm eff}(x)}} \ .
\label{eq:effav}
\eeq

\subsection{The replicon}
\label{app:replicon}

We also need to determine the transition line to the RSB phase. This is done by calculating the \emph{replicon} eigenvalue of the matrix of second derivatives of the replica action~\cite{parisiSGandbeyond}. The replica action is
\beq
S(Q_{ab}) = \frac{(\beta J)^2}{4}\sum_{a b}^n Q^2_{ab} - \log\left[\overline{\int \dd^n x\ \exp\left(-\frac{\beta\kappa}{2}\sumn x_a^2 - \frac{\beta }{4!}\sumn x_a^4 + \beta h\sumn x_a + \frac{(\beta J)^2}{2}\sum_{ab}^n x_aQ_{ab}x_b\right) }\right],
\eeq
where the overline denotes an average over $p(\kappa)$. We wish to calculate the tensor of second derivatives of this action with respect to $Q_{ab}$,
\beq
M_{ab;cd} \equiv \frac{\partial S}{\partial Q_{ab}\partial Q_{cd}} = M_1\left(\frac{\d_{ac}\d_{bd} + \d_{ad}\d_{bc}}{2}\right) + M_2\left(\frac{\d_{ac} + \d_{bd} + \d_{ad} + \d_{bc}}{4}\right) +M_3,
\label{eq:hessian_app}
\eeq
where $\d_{ij}$ is simply a Kronecker delta, and the last expression is the most general form that can be taken by a replica-symmetric tensor with four indices
(here grouped as $ab;cd$ to emphasize that the first two indices are related to the first derivative with respect to $Q_{ab}$, and the other two to the second derivative)~\cite{SimpleGlasses2020}. We recall that the replicon mode is simply given by~\cite{SimpleGlasses2020}
\beq
\l_{\rm R}  = M_1.
\eeq
The derivatives of the first (kinetic) term are easy to take, and one easily gets
\beq
\frac{\partial S_{\rm kin}}{\partial Q_{ab}\partial Q_{cd}} = (\beta J)^2 \left(\frac{\d_{ac}\d_{bd} + \d_{ad}\d_{bc}}{2}\right).
\label{eq:hesskin}
\eeq
The derivatives of the second (interaction) term are more cumbersome. But using a compact notation, one can write them down as
\beq
\frac{\partial S_{\rm pot}}{\partial Q_{ab}\partial Q_{cd}} = \frac{(\beta J)^4}{2} \left[\thavn{x_a x_b x_c x_d} - \thavn{x_a x_b}\thavn{x_cx_d} \right],
\eeq
with the $\thavn{\bullet}$ averages defined, for a replica-symmetric $Q_{ab}$, as
\beq
\thavn{\mathcal{O}(x_1,\dots,x_n)} \equiv \frac{1}{\overline{(Z_{\rm eff})^n}}\overline{\int \dd^nx\ \mathcal{O}(x_1,\dots,x_n) \prod_{i=1}^n e^{-\beta v_{\rm eff}(x_a)}},
\eeq
and $v_{\rm eff}(x)$ has been defined in Eq.~\eqref{eq:veff_app}. Now the overline also indicates an average over the Gaussian measure $\DD z$.

In order to calculate $M_1$ (and therefore the replicon), we can just observe that 
\beq
M_1 = 2M_{12;12} - 4M_{12;13} + 2M_{12;34},
\eeq
which comes from direct inspection of Eq.~\eqref{eq:hessian_app}. The kinetic part is trivially obtained from Eq.~\eqref{eq:hesskin}
\beq
M_1^{\rm kin} = (\beta J)^2 \ ,
\eeq
while, for the interaction part, we need to compute the averages
\beq
\begin{split}
M_1^{\rm int} = \frac{(\beta J)^4}{2} &\left[2(\thavn{x_1x_2x_1x_2} - \thavn{x_1x_2}\thavn{x_1x_2})-4(\thavn{x_1x_2x_1x_3} - \thavn{x_1x_2}\thavn{x_1x_3})\right.\\
&\left.+2(\thavn{x_1x_2x_3x_4} - \thavn{x_1x_2}\thavn{x_3x_4})\right],
\end{split}
\eeq
at RS level, and for $n\to 0$. Thanks to replica symmetry, one has
\beq
\thavn{x_a x_b} = \thav{x_cx_d}_n,\ \forall a,b,c,d:\ a\neq b,\ c\neq d,
\eeq
so the expression for the replicon reduces to
\beq
M_1^{\rm int} = \frac{(\beta J)^4}{2} \left[2\thavn{x_1x_2x_1x_2} -4\thavn{x_1x_2x_1x_3} +2\thavn{x_1x_2x_3x_4}\right].
\eeq
We now need to calculate these averages for $n\to 0$. The first one is
\beq
\lim_{n\to 0 } \thavn{x_1x_2x_1x_2} = \lim_{n\to 0} \frac{1}{\overline{(Z_{\rm eff})^n}} \overline{\left(\int \dd x_1\  x_1^2e^{-\beta v_{\rm eff}(x_1)}\right)\left(\int \dd x_2\  x_2^2e^{-\beta v_{\rm eff}(x_2)}\right)\left(\int \dd x\  e^{-\beta v_{\rm eff}(x)}\right)^{n-2}} = \overline {\thav{x^2}^2},
\eeq
where the $\thav{\bullet}$ average is defined as in Eq.~\eqref{eq:effav}. For the second term, we have
\beq
\begin{split}
\lim_{n\to 0 } \thavn{x_1x_2x_1x_3} =\ &\lim_{n\to 0} \frac{1}{\overline{(Z_{\rm eff})^n}} \overline{\left(\int \dd x_1\  x_1^2e^{-\beta v_{\rm eff}(x_1)}\right)\left(\int \dd x_2\  x_2e^{-\beta v_{\rm eff}(x_2)}\right)}\\
& \overline{\times \left(\int \dd x_3\  x_3e^{-\beta v_{\rm eff}(x_3)}\right)\left(\int \dd x\  e^{-\beta v_{\rm eff}(x)}\right)^{n-3}}=\overline{\thav{x^2} \thav{x}^2} \ ,
\end{split}
\eeq
and for the third, one can easily get by the same logic
\beq
\lim_{n\to 0 } \thavn{x_1x_2x_3x_4} = \overline{\thav{x}^4} \ .
\eeq
In summary, one has for $M_1^{\rm int}$
\beq
 M^{\rm int}_1 = \frac{(\beta J)^4}{2} [2\overline {\thav{x^2}^2} - 4\overline{\thav{x^2} \thav{x}^2} + 2 \overline{\thav{x}^4}] = (\beta J)^4 \overline{(\thav{x^2} - \thav{x}^2)^2},
\eeq
and the final expression of the replicon eigenvalue, factoring out a positive constant, reads:
\beq
\lambda_R \propto 1 - (\beta J)^2 \overline{(\thav{x^2} - \thav{x}^2)^2} \ .
\label{eq:replicon_app}
\eeq

\subsection{The $T\to0$ limit \label{subsec:zeroT}}
\label{app:Tzero}

In order to compute the spectrum, we need to find the ground state of the system in the athermal limit. In that limit, one can 
easily see that $q \to \qt$ linearly in $T$, so we 
define the following ``athermal'' overlaps and their associated saddle point equations,
\beq
\begin{split}
\chi=\beta(\qt - q) &= \int  \dd p(\kappa) \int \DD z \frac{z\left<x\right>}{\sqrt{\qt} J} \ ,\\
\qt &= \int \dd p(\kappa) \int \DD z \left<x^2\right> \ .
\label{eq:chi}
\end{split}
\eeq
In the zero temperature limit, the equilibrium averages $\thav{x}$ on the effective potential $v_{\rm eff}(x)$ become dominated by its ground state. One has then
\beq
\lim_{T\to 0}\thav{x} = x^*(z,\k),
\eeq
where $x^*$ is the absolute minimum of the potential in Eq.~\eqref{eq:veff_app}, which in this limit reads
\beq
v_{\rm eff}(x) \equiv \frac{m}{2}x^2 + \frac{1}{4!}x^4 - (f+h)x \ ,
\label{eq:veff_zero_T_app}
\eeq
with the definitions $m\equiv \kappa-J^2\chi$ and $f\equiv Jz\sqrt{\qt}$, as given in the main text. The Eqs.~(\ref{eq:chi}) can then be written as follows
\beq
\begin{split}
\chi &= \int  \dd p(m) \int \dd p(f) \frac{fx^*(f,m)}{J^2\qt} \ ,\\
\qt &= \int \dd p(m) \int \dd p(f) \left(x^*(f,m)\right)^2 ,
\label{eq:sp_zero_T_app}
\end{split}
\eeq
where 
\beq
p(f)\equiv\NN(0,J^2\qt) \ , \qquad p(m) = U(\kappa_m-J^2\chi,\kappa_{_{\!M}}-J^2\chi) \ .
\eeq
To solve them, one can proceed as follows. Starting from a guess for $\chi$ and $\qt$, one first generates the two random parameters $(z,\k)$, and for each realization, one finds the minimum of the effective potential, by solving the cubic equation 
 \beq
 v'_{\rm eff}(x)=0 \ .
 \eeq
Because the potential is quartic, an analytical solution of the cubic equation can be obtained and is given explicitly in Appendix~\ref{sec:phasediag}.
One then averages over the random variables $(z,\k)$ to compute the r.h.s. of Eqs.~\eqref{eq:sp_zero_T_app} and obtain new estimates of $\chi$ and $\qt$.
The procedure is iterated until convergence.
In Appendix~\ref{sec:phasediag} we provide the detailed algorithms we used to obtain the phase diagram reported in the main text. 

We note that an alternative equation for $\chi$, which we report in the main text and is more useful when it comes to understating the location $\lambda_m$ of the spectrum's lower edge, can be obtained. We start from the first of Eqs.~\eqref{eq:chi}, at finite temperature, and we rewrite it as
\beq
\begin{split}
\chi =\ &\int  \dd p(\kappa) \int \DD z \frac{z }{\b J^2 \qt} \frac{\dd}{\dd z}  \log \int \dd x e^{-\b v_{\rm eff}(x)}
=\int  \dd p(\kappa) \int \DD z \frac{1 }{\b J^2 \qt} \frac{\dd^2}{\dd z^2}  \log \int \dd x e^{-\b v_{\rm eff}(x)} \\
=&\int  \dd p(m) \int \dd p(f) \frac{1 }{\b} \frac{\dd^2}{\dd f^2}  \log \int \dd x e^{-\b v_{\rm eff}(x)}
=\int  \dd p(m) \int \dd p(f) \frac{\dd\left<x\right>}{\dd f}  \ ,
\label{eq:chigen}
\end{split}
\eeq
where we used the following relation, easily obtained by integration by parts and valid for any function $g(z)$:
\beq
\int \DD z z  g'(z) = \int \DD z g''(z) \ .
\eeq
The $T\to 0 $ limit of this expression needs to be taken carefully, as $\langle x\rangle \to x^*$ (the absolute minimum of the effective potential) in that limit, and $x^*$ is not guaranteed to be a smooth function of $f$. In fact, if the effective potential $v_{\rm eff}(x;f,m)$ has multiple minima (i.e.  it is a double well), then $x^*$ will jump discontinuously when the sign of the linear term $f+h$ changes, as the absolute minimum switches from one side of the origin to the other. This will happen as soon as $m_m<0$ as discussed in the main text. 
Away from the singularity, because $x^*(f)$ is the solution of $v'_{\rm eff}(x^*,f)=0$, one has
\beq
0 = \frac{\dd}{\dd f} v'_{\rm eff}[x^*(f),f)] = v''_{\rm eff}[x^*(f),f)] \frac{\dd x^*}{ \dd f  } - 1
\qquad \Rightarrow \qquad
\frac{\dd x^*}{ \dd f  } = \frac{1}{v''_{\rm eff}(x^*)}  \ . 
\eeq
Adding the singular term, the proper limit of $\frac{\dd\left<x\right>}{\dd f}$ therefore is
\beq
\frac{\dd x^*}{ \dd f  } = \frac{1}{v''_{\rm eff}(x^*)} + [x^*(-h^+) - x^*(-h^-)] \d(f+h) \ .
\label{eq:dxstardf}
\eeq
As long as $m_m\geq 0$, no DW are present, the second term vanishes and one has the equation for $\chi$
\beq\label{eq:chiedge}
\chi =   \int  \dd p(m) \int \dd p(f)  \frac{1}{v_{\rm eff}''(x^*(f,m))}
= \overline{\left[\frac{1}{v_{\rm eff}''(x^*(f,m))}\right]}
 \ ,
\eeq
i.e., in absence of DWs, $\chi$ is the average of the inverse of the curvature of the effective potential in its minimum. This is the equation that we report in the main text and we use to prove that $\lambda_m=0$ on the RSB transition line. Note that the overline denotes the average over $f,m$ which is indicated as $\langle\bullet\rangle_{f,m}$ in the main text.

The last ingredient we miss is the $T\to0$ limit of the replicon, Eq.~\eqref{eq:replicon_app}. Using the definition in Eq.~\eqref{eq:effav}, one can write
\beq
(\beta J)^2\overline{(\thav{x^2}-\thav{x}^2)^2} = J^2 \overline{\left(\frac{\dd \left<x\right>}{\dd f}\right)^2} \ .
\eeq
Therefore one has, for $T\to 0$
\beq
\lambda_R = 1 - J^2\overline{\left(\frac{\dd x^*}{\dd f}\right)^2} \ .
\eeq
Notice that then, when $m_m<0$ and DWs are present, Eq.~\eqref{eq:dxstardf} implies that the replicon is the average of the square of a delta function, which then formally diverges to $-\infty$, hence the replica symmetry is automatically broken. As we state in the main text, the presence of DWs in the ensemble of effective potentials is a sufficient condition for a RSB glass transition to take place in our model, with the replicon jumping to minus infinity rather than vanishing. On the contrary, when $m_m>0$, the second term in Eq.~\eqref{eq:dxstardf} vanishes and one can simply write for the replicon
\beq
\l_R = 1 - J^2\overline{\left[\frac{1}{v_{\rm eff}''(x^*(f,m))^2}\right]}.
\label{eq:replicon_zero_T_app}
\eeq
which is the expression given and used in the main text, valid in the RS phase and in absence of double wells.

\section{Analysis of the effective potential}


We now focus on the effective potential. In particular, we want to compute the statistics of the diagonal elements $a\equiv m+J^2\chi + x^2/2$.

\subsection{Effective potential and ground state}
\label{sec:effpot}

The effective potential has the form
\beq\begin{split}
v_{\rm eff}(x) &=  \frac{1}{4! }x^4 + \frac{m}2 x^2 - H x \ ,  \qquad
v_{\rm eff}'(x) =  \frac{1}{3! }x^3 + m x   - H  \ , \qquad
v_{\rm eff}''(x) =  \frac{1}{2 }x^2 + m   \ ,
\end{split}\eeq
with the two effective parameters
\beq\begin{split}
&m = \k - J^2\c \ , \qquad\quad  p(m) = U(\kappa_m-J^2\c, \kappa_{_{\!M}}-J^2\c)  \ , \\
&H =  h + f \ , \qquad\qquad p(H) = \NN(h, J^2\tilde q) \ .
\end{split}\eeq
The equation for the stationary points of the effective potential is a depressed cubic~\footnote{\url{https://en.wikipedia.org/wiki/Cubic_equation}} of the form
\beq
v_{\rm eff}'(x) =  \frac{1}{6}\left( x^3 + P x  +Q \right)= 0 \ , \qquad P = 6m  \ , \qquad Q = -6H  \ ,
\eeq
whose discriminant is
\beq
\D = 4 P^3 + 27 Q^2 \ \ \propto \ \ m^3 + \frac{9}8 H^2  \ .
\eeq
Hence the solutions are organised as follows:
\begin{itemize}
\item
For $\D<0$ there are three real solutions:
\beq
x_k = 2 \sqrt{\frac{-P}3 } \cos\left[
\frac13 \text{arccos}\left(
\frac{3Q}{2P}\sqrt{\frac{-3}P }
\right) - \frac{2\pi k}{3}
\right] \ , \qquad k=0,1,2 \ ,
\eeq
Note that $\D<0$ implies $P<0$ and the solution corresponding to the absolute minimum can be written as
\beq
x^* = -2\text{sgn}(Q) \sqrt{\frac{|P|}3 } \cos\left[
\frac13 \text{arccos}\left(
\frac{3|Q|}{2|P|}\sqrt{\frac{3}{|P|} }
\right) 
\right] \ .
\eeq
\item
For $\D>0$ there a single real solution:
\beq
x^*= \begin{cases}
-2 \text{sgn}(Q) \sqrt{\frac{-P}3 } \cosh\left[
\frac13 \text{arccosh}\left(
\frac{-3|Q|}{2P}\sqrt{\frac{-3}P }
\right) \right] & \text{ for } P<0  \ ,\\
-2 \sqrt{\frac{P}3 } \sinh\left[
\frac13 \text{arcsinh}\left(
\frac{3Q}{2P}\sqrt{\frac{3}P }
\right) 
\right]& \text{ for } P>0 \ . 
\end{cases}
\eeq
\item
For $\D=0$ there are two possibilities:
\begin{itemize}
\item $P=Q=0$ and $x=0$ is a triple root, i.e. $v_{\rm eff}(x) =  x^4/4!$;
\item $P\neq 0$ and then $x=3Q/P$ is a single root and $x=-3Q/(2P)$ is a double root.
\end{itemize}
\end{itemize}
To summarize, the ground state can be written as follows:
\beq\label{eq:GSeffpot}
x^*(m,H) = 2 \text{sgn}(H) \sqrt{2 |m|} \FF_{\text{sgn}(m)}\left(\frac{3 |H|}{(2 |m|)^{3/2}} \right) \ ,
\qquad H=h+f \ , \qquad m = \k - J^2 \c \ ,
\eeq
with
\beq
\FF_+(\xi)=\sinh\left[\frac13 \text{arcsinh} \xi\right] \ , \qquad
\FF_-(\xi) = \begin{cases}
\cos\left[\frac13 \text{arccos} \xi\right] & \xi<1 \ , \\
\cosh\left[\frac13 \text{arccosh} \xi\right] & \xi>1 \ .
\end{cases}
\eeq

\subsection{Double wells and distribution of curvatures\label{subsec:pofa}}
\label{sec:DW}

Using the auxiliary formula
\beq
\int_{-\ee}^\ee \dd p(H) = \frac12 \left[ \text{erf}\left( \frac{h+\ee}{J \sqrt{2 \tilde q}} \right)-\text{erf}\left( \frac{h-\ee}{J \sqrt{2 \tilde q}} \right) \right]
\approx 2 \ee \frac{e^{-\frac{h^2}{2 J^2 \tilde q}}}{\sqrt{2\pi J^2 \tilde q}} \ , \qquad \ee \ll J \sqrt{\tilde q}  \ ,
\eeq
we get as a first result the fraction of double well potentials in the ensemble, which is given by:
\beq
p_{\rm dw} = p(\D<0) =\int_{\k_m-J^2 \c}^{\min\{0,\k_{_{\!M}}-J^2\c\}} \frac{\dd m}{\k_{_{\!M}}-\k_m} \int_{-\sqrt{-8 m^3 /9}}^{\sqrt{-8 m^3 /9}} \dd p(H)  \ .
\eeq
In particular, when $\k_{_{\!M}} > 0$ and $h>0$,  and $k_m-J^2 \c \to 0^-$, we have
\beq
p_{\rm dw} \approx \frac{8\sqrt{2}}{15} \frac{(J^2\chi-\kappa_{m})^{5/2}}{\k_{_{\!M}}-\k_m}p(0) \ ,
\qquad
p(0)= \frac{e^{-\frac{h^2}{2 J^2 \tilde q}}}{\sqrt{2\pi J^2 \tilde q}} \ .
\eeq

\begin{figure}[t]
\begin{center}
\includegraphics[width=0.65\textwidth]{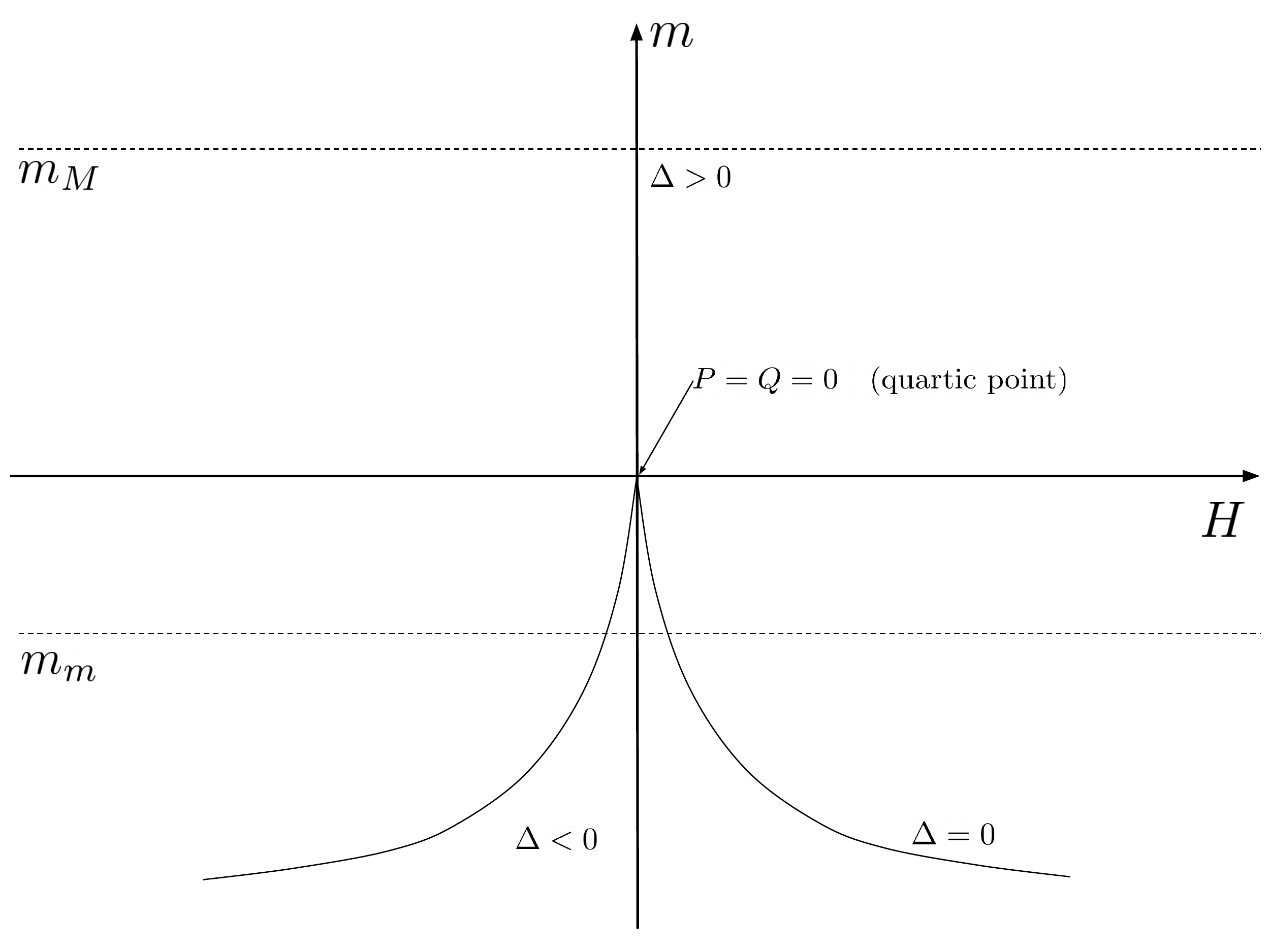}\nolinebreak
\caption{Sketch of the integration region for $p(\tilde a)$.}
\label{fig:a}
\end{center}
\end{figure}

The second result  is the distribution of the curvatures $\tilde{a}$ in the ground state, related by a simple shift to the distribution of diagonal elements $a$,
\beq
\tilde a = v''_{\rm eff}(x^*) = m + \frac{(x^*)^2}2 = a -J^2 \c  \ .
\eeq
We are interested in the small $\tilde a$ behavior, which is obtained following similar steps as in 
the soft potential model analysis~\cite{soft_potential_model_1991,Schober_prb_1992,GC03}. 
First of all, we note that the distribution of curvatures can be either gapped or gapless (the curvature cannot be negative). The only possibility to have $\tilde a=0$ (i.e., a quartic potential) is to have $P=Q=0$, or equivalently $m=H=0$. Hence, if $m_m > 0$ or $m_{_{\!M}}<0$, the distribution
of $\tilde a$ is gapped.

We then assume that $m_m=\kappa_m-J^2 \chi \leq 0$ and $m_{M} = \kappa_{_{\!M}}-J^2\c \geq 0$; in this case, the distribution of $\tilde{a}$ is gapless, which implies 
$a_m=J^2\chi$ and $\lambda_m=0$, as we state in the main text. We shall focus on this particular case, which is the one relevant for the $\omega^4$-transition line.
In this case, the integration domain over 
the random variables $m$ and $H$ can be decomposed as sketched in Fig.~\ref{fig:a}. 
For $m_{_{\!M}}>0$,
the contribution of positive $m$ to the cumulative distribution can be written, when $\tilde a\to 0$, as
\beq\begin{split}
G_+(\tilde a) &= \int_0^{\tilde a} \dd m \int \dd p(H) \th[m + \frac12 x^2  < \tilde a ] \sim p(0) \tilde a^{5/2} \int \dd\eta\int_0^1 \dd\xi \xi^{3/2}  
\th[\xi(1 + \frac12 y^2)  < 1 ] \\
&=p(0) \frac{2}{5} \tilde a^{5/2}  \int \dd\eta \left(1 + \frac12 y^2\right)^{-5/2}  = 1.13137\ldots \times p(0) \tilde a^{5/2} 
\ ,
\end{split}\eeq
where we introduced $\xi = m/\tilde a$, $\eta = H / (\xi \tilde a)^{3/2}$ and 
\beq
y = \frac{x}{ \sqrt{\xi \tilde a}} = -2 \sqrt{2 } \sinh\left[
\frac13 \text{arcsinh}\left(
- \frac{3 \h}{2 \sqrt{2}}
\right) 
\right] \ .
\eeq
The contribution of negative $m$ with $\D>0$ can be written, by similar means, as
\beq\begin{split}
G_-^1(\tilde a) &= \int_{m<0} \dd m \int \dd p(H) \th[m + \frac12 x^2  < \tilde a ]\th(\D>0) \sim p(0) \tilde a^{5/2} \int_{\h^2>8/9} \dd\eta \int_{\xi>0} \dd\xi \xi^{3/2}  
\th[\xi(-1 + \frac12 y^2)  < 1 ] \\
&=p(0) \frac{2}{5} \tilde a^{5/2}  \int_{\h^2>8/9} \dd\eta \left(-1 + \frac12 y^2\right)^{-5/2}  = 0.175024\ldots \times p(0) \tilde a^{5/2} 
\ , \\
y &=  2 \sqrt{2 } \text{sgn}(\h) \cosh\left[
\frac13 \text{arccosh}\left(
\frac{3 |\h|}{2 \sqrt{2}}
\right) 
\right] \ .
\end{split}\eeq
Finally, the contribution of negative $m$ with $\D<0$ is
\beq\begin{split}
G_-^3(\tilde a) &= \int_{m<0} \dd m \int \dd p(H) \th[m + \frac12 x^2  < \tilde a ]\th(\D<0) \sim p(0) \tilde a^{5/2} \int_{\h^2<8/9} \dd\eta \int_{\xi>0} \dd\xi \xi^{3/2}  
\th[\xi(-1 + \frac12 y^2)  < 1 ] \\
&=p(0) \frac{2}{5} \tilde a^{5/2}  \int_{\h^2<8/9} \dd\eta \left(-1 + \frac12 y^2\right)^{-5/2}  = 0.0792461\ldots \times p(0) \tilde a^{5/2} 
\ ,\\
y &=  2 \sqrt{2 } \text{sgn}(\h) \cos\left[
\frac13 \text{arccos}\left(
\frac{3 |\h|}{2 \sqrt{2}}
\right) 
\right] \ .
\end{split}\eeq
Collecting these results, we obtain
\beq\begin{split}
G(\tilde a)&=  1.3856401\ldots \times \frac{e^{-\frac{ h^2}{2J^2\tilde q}}}{\sqrt{2 \pi J^2 \tilde q}} \tilde a^{5/2}\ , \qquad
\text{for}\qquad \kappa_m<J^2\chi<\kappa_{_{\!M}} \ ,\\
p(\tilde a)&= \AA_d \tilde a^{3/2} \ , \qquad \AA_d(J) = 3.46410025\ldots \times \frac{e^{-\frac{ h^2}{2J^2\tilde q}}}{\sqrt{2 \pi J^2 \tilde q}}\ , \qquad
\text{for}\qquad \kappa_m<J^2\chi<\kappa_{_{\!M}} \ .
\end{split}\eeq
This applies whenever $m_m< 0$ and $m_{_{\!M}}>0$, and implies $p(a) \sim \mathcal{A}_d(a-J^2\chi)^{3/2}$, which leads to the results of the main text in terms of location and shape of the spectrum edge. Note that if $(m_m=0,m_{_{\!M}}>0)$ or $(m_m<0,m_{_{\!M}}=0)$, one also obtains the $\tilde a^{3/2}$ law, but with a different prefactor because the contribution of positive (or negative) $m$ is absent.

\section{Drawing the phase diagram\label{sec:phasediag}}

In this section we report the algorithm used to determine the $\omega^2$- and $\omega^4$-transition lines in the phase diagram, building up from the equations derived in the previous section. We place ourselves in the RS region of the phase diagram, with the aim of determining its boundaries. We start from the form of the RS Eqs.~\eqref{eq:sp_zero_T_app}  for $\tilde{q}$ and $\chi$, with $x^*(z,m)$ being the \emph{unique} ground state of the effective potential (having double wells would automatically imply RSB, as detailed in Appendix~\ref{subsec:zeroT} and the main text), given by Eq.~\eqref{eq:GSeffpot}.
Because we also want to explore the limits $J\to 0$ and $h\to 0$ (which implies $\tilde{q}\to0$ in the RS phase), it is convenient to perform the rescaling
\beq
J^2 \tilde q \to \tilde q,
\eeq
so that the equations take the form
\beq
\begin{split}
\tilde q&=J^2\int_{m_m}^{m_{_{\!M}}}\dd p(m) \int_{-\infty}^\infty \frac{\dd z}{\sqrt{2\pi}} e^{-z^2/2} \left(x^*(z,m)\right)^2 \ ,\\
\chi &= \int_{m_m}^{m_{_{\!M}}} \dd p(m) \int_{-\infty}^\infty \frac{\dd z}{\sqrt{2\pi}} e^{-z^2/2}\frac{z x^*(z,m)}{\sqrt{\tilde q}} \ ,
\end{split}
\label{RS_scaled}
\eeq
and $x^*$ is now the unique solution of the equation 
\beq
\frac{x^3}{6} + m x - h + \sqrt{\tilde q}z=0 \ ,
\label{cubic}
\eeq
given by an expression similar to Eq.~\eqref{eq:GSeffpot}.
Furthermore, the equation for $\chi$ can be rewritten in the following way
\beq
\chi = \int_{m_m}^{m_{_{\!M}}}\dd p(m) \int_{-\infty}^\infty \frac{\dd z}{\sqrt{2\pi}} e^{-z^2/2}\frac{1}{m+\frac 12\left(x_*(z,m)\right)^2} \ ,
\label{better_chi}
\eeq
which is the rescaled form of Eq.~\eqref{eq:chiedge} and completely equivalent to its form in Eqs.~\eqref{RS_scaled} everywhere in the RS phase. However, we found that this form is better behaved under numerical resolution.

There are two ways to break the replica symmetry at $T=0$, as discussed in 
the main text:
\begin{itemize}
\item[(i)] The replicon vanishes continuously, $\l_R=0$, but the minimal effective stiffness stays positive, $m_m>0$. This case corresponds to having a GOE-like spectrum, with an $\omega^2$ low-frequency tail populated by delocalized modes. This is a standard RSB transition.
\item[(ii)] The minimal effective stiffness vanishes, $m_m = \k_m-J^2\chi = 0$ with $\l_R>0$.  At this point, DW effective potentials appear. The replica symmetry is then broken via a discontinuity in replicon eigenvalue, which jumps to $-\infty$ beyond the transition. This case corresponds to having a DIAG-like spectrum, with a $\omega^4$ low-frequency tail and partially localized modes near the edge.
\end{itemize}

\subsection{Finding the transition lines}

\subsubsection{$\omega^4$-transition} 

On the $\omega^4$-transition line one has 
$m_m=0$.
Therefore one can take Eqs.~\eqref{RS_scaled}, set $m_m=0$,
\beq
\begin{split}
\tilde q&=J^2\int_{0}^{m_{_{\!M}}}\dd p(m) \int_{-\infty}^\infty \frac{\dd z}{\sqrt{2\pi}} e^{-z^2/2} \left(x_*(z,m)\right)^2 \ ,\\
\k_m &= J^2\int_{0}^{m_{_{\!M}}}\dd p(m) \int_{-\infty}^\infty \frac{\dd z}{\sqrt{2\pi}} e^{-z^2/2}\frac{1}{m+\frac 12\left(x_*(z,m)\right)^2} \ ,
\end{split}
\label{o4}
\eeq
and solve them to find $h_c(J)$ and $\tilde q$ at fixed $J$. $h_c(J)$ is the critical line in this case, and we remind that $\kappa_m$ is a fixed model parameter. This can be achieved via the following numerical scheme: 
\begin{algorithm}[H]
	\SetAlgoLined
	fix $J$\;
	initialize $\tilde q$ and $h$\;
	\While{$\tilde q$ and $h$ not converging}{
	$\tilde q \gets ({\rm damped})J^2\int_{0}^{m_{_{\!M}}}\dd p(m) \int_{-\infty}^\infty \frac{\dd z}{\sqrt{2\pi}} e^{-z^2/2} \left(x^*(z,m)\right)^2$\;
	$h \gets ({\rm damped})(h-\k_m+J^2\int_{0}^{m_{_{\!M}}} \dd p(m) \int_{-\infty}^\infty \frac{\dd z}{\sqrt{2\pi}} e^{-z^2/2}\frac{1}{m+\frac 12\left(x^*(z,m)\right)^2})$ \; 
	}
	\KwResult{$(\tilde q,h=h_c(J))$ are the values of the corresponding parameters at the transition point, for each $J$.}
	\caption{$\omega^4$-transition line}
\end{algorithm}
In order to ensure that the transition is $\omega^4$-like, one needs to prove that $\l_R>0$ at the transition. The expression for $\l_R$ however contains integrable singularities that could make its numerical computation unstable. We derive below an expression for $\l_R$ that does not suffer from these problems, and furthermore proves that $\l_R$ is indeed positive at the transition.

\subsubsection{$\omega^2$-transition}
In this case, one has $\l_R=0$
at the transition,
but differently from the previous case one still needs to determine both $\tilde q$ and $\chi$ trough Eqs.~\eqref{RS_scaled}, and only then get the transition point from the ${\l_R=0}$ condition. Therefore we need to find also $J_c(h)$ at fixed $h$. A slightly more complicated numerical scheme, which we report below, is needed (note the update for $J_c$ which avoids bisection methods):
\begin{algorithm}[H]
	\SetAlgoLined
	fix $h$\;
	initialize $\tilde q$, $\chi$ and $J_c$\;
	\While{$\tilde q$, $\chi$ and $J_c$ not converging}{
	$\tilde q \gets ({\rm damped})J_c^2\int_{m_m}^{m_{_{\!M}}}\dd p(m) \int_{-\infty}^\infty \frac{\dd z}{\sqrt{2\pi}} e^{-z^2/2} \left(x^*(z,m)\right)^2$\;
	$\chi \gets ({\rm damped})\left(J_c^2\int_{m_m}^{m_{_{\!M}}}\dd p(m) \int_{-\infty}^\infty \frac{\dd z}{\sqrt{2\pi}} e^{-z^2/2}\frac{1}{m+\frac 12\left(x^*(z,m)\right)^2}\right)$ \; 
	$J_c\gets ({\rm damped})\left(J_c+\left[1-J_c^2\int_{m_m}^{m_{_{\!M}}}\dd p(m) \int_{-\infty}^\infty \frac{\dd z}{\sqrt{2\pi}} e^{-z^2/2}\frac{1}{\left[m+\frac 12\left(x^*(z,m)\right)^2\right]^2}\right]\right)$
	}
	\KwResult{$(\tilde q,\chi,J_c(h))$ are the values of the corresponding parameters at the transition point.}
	\caption{$\omega^2$-transition line}
\end{algorithm}

\subsection{Numerically stable expression for the replicon}
We recall the expression in Eq.~\eqref{eq:replicon_zero_T_app} of the replicon eigenvalue in the RS phase, in explicit form:
\beq
\l_R= 1-J^2\int_{m_m}^{m_{_{\!M}}}\dd p(m) \int_{-\infty}^\infty \frac{\dd z}{\sqrt{2\pi}} e^{-z^2/2}\frac{1}{\left[m+\frac 12\left(x_*(z,m)\right)^2\right]^2} \ .
\eeq
We remark that the denominator appearing in the integral is essentially $\tilde{a}^2$. The expression could then be equivalently rewritten as 
\beq
\begin{split}
\l_R &= 1-J^2\int \dd \tilde{a} \int_{m_m}^{m_{_{\!M}}}\dd p(m) \int_{-\infty}^\infty \frac{\dd z}{\sqrt{2\pi}} e^{-z^2/2}\frac{1}{\left[m+\frac 12\left(x_*(z,m)\right)^2\right]^2}\delta\left(\tilde{a} - m -  \frac 12\left(x_*(z,m)\right)^2\right) \\ &= 1-J^2\int \dd \tilde{a} \frac{p(\tilde{a})}{\tilde{a}^2} \ .
\end{split}
\eeq
At the transition line, the distribution of $\tilde{a}$ is gapless and follows $p(\tilde{a})\sim \tilde{a}^{3/2}$ near its edge, as discussed in section~\ref{subsec:pofa}.
The above expression above highlights the singularity of the integrand at $\tilde{a}=0$; this singularity is integrable, which proves that $\l_R$ is finite at the transition, and that the jump to $-\infty$ is due to the singular $\delta(x+h)$ term in Eq.~\eqref{eq:dxstardf}, while the $\frac{1}{v_{\rm eff}''(x^*)}$ term always stays finite and positive. Still, the singularity could cause problems when evaluating the replicon numerically. Is is possible to manipulate this expression to obtain an alternative one that, while being more cumbersome, contains no singularities. Using the fact that in the RS phase $m_m\geq 0$ and $x^*(z,m)$ is the \emph{unique} solution of Eq.~\eqref{cubic}, we can rewrite $\II_R=\int \dd \tilde{a} p(\tilde{a})/\tilde{a}^2$ as
\beq
\begin{split}
\II_R&=\int_{-\infty}^\infty \dd x\int_{m_m}^{m_{_{\!M}}}\dd p(m) \int_{-\infty}^\infty \frac{\dd z}{\sqrt{2\pi}} e^{-z^2/2}\frac{1}{m+\frac 12x^2}\delta\left(\frac{x^3}{6} + mx -h +\sqrt{\tilde q}z\right)\\
&=\int_{-\infty}^\infty \dd x\int_{-\infty}^\infty\frac{\dd \hat x}{2\pi}\int_{m_m}^{m_{_{\!M}}}\dd p(m) \int_{-\infty}^\infty \frac{\dd z}{\sqrt{2\pi}} e^{-z^2/2}\frac{1}{m+\frac 12x^2}\exp\left[i\hat x\left(\frac{x^3}{6} + mx -h +\sqrt{\tilde q}z\right)\right]\\
&=\int_{-\infty}^\infty\frac {\dd x}{\sqrt{2\pi\tilde q}}\int_{m_m+x^2/2}^{m_{_{\!M}}+x^2/2}\dd p(m)\frac{1}{m}\exp\left[-\frac{1}{2\tilde q}\left(\frac{x^3}{6} + \left(m-\frac{x^2}{2}\right)x -h \right)^2\right] \ .
\end{split}
\eeq
If $m_m=0$, this expression has an integrable singularity at $m=0$, which we can eliminate with an integration by parts. In doing so we obtain
\beq
\begin{split}
\II_R&=\II_0+\II_1+\II_2 \ , \\
\II_0&=\frac{1}{\D_\kappa}\int_{-\infty}^\infty\frac {\dd x}{\sqrt{2\pi\tilde q}} \ln\left(m_{_{\!M}}+\frac{x^2}{2}\right)\exp\left[-\frac{1}{2\tilde q}\left(\frac{x^3}{6} + m_{_{\!M}}x -h \right)^2\right] \ ,\\
\II_1&=-\frac{1}{\D_\kappa}\int_{-\infty}^\infty\frac {\dd x}{\sqrt{2\pi\tilde q}} \ln\left(m_m+\frac{x^2}{2}\right)\exp\left[-\frac{1}{2\tilde q}\left(\frac{x^3}{6} + m_mx -h \right)^2\right] \ ,\\
\II_2&=-\int_{-\infty}^\infty\frac {\dd x}{\sqrt{2\pi\tilde q}}\int_{m_m+x^2/2}^{m_{_{\!M}}+x^2/2}\frac{\dd m}{\D_\kappa} \left(\ln m\right) \frac{\dd}{\dd m}\exp\left[-\frac{1}{2\tilde q}\left(\frac{x^3}{6} + \left(m-\frac{x^2}{2}\right)x -h \right)^2\right] \ ,
\end{split}
\eeq
where $\Delta_\kappa\equiv \kappa_{_{\!M}}-\kappa_m$. The first integral $\II_0$ is perfectly convergent assuming $m_{_{\!M}}>0$. The second integral $\II_1$, however, has some integrable singularity if $m_m=0$. By assuming $m_m=0$, we can rewrite the integral as
\beq
\II_1= \frac{1}{\D_\kappa}\int_{-\infty}^\infty\frac {\dd x}{\sqrt{2\pi\tilde q}} \left[\ln 2 +2\left(x\ln |x|-x\right)\frac{\dd}{\dd x} \right]\exp\left[-\frac{1}{2\tilde q}\left(\frac{x^3}{6} -h \right)^2\right],
\eeq
which now contains no singularities. Finally we need to consider $\II_2$. Integrating again the logarithmic singularity by parts, one obtains
\beq
\begin{split}
\II_2&= -\frac{1}{\D_\kappa}\int_{-\infty}^\infty\frac {\dd x}{\sqrt{2\pi\tilde q}} \left[\left(m_{_{\!M}}+\frac{x^2}{2}\right)\ln \left(m_{_{\!M}}+\frac{x^2}{2}\right) - \left(m_{_{\!M}}+\frac{x^2}{2}\right)\right] \left[\left.\frac{\dd}{\dd m}\exp\left[-\frac{1}{2\tilde q}\left(\frac{x^3}{6} + mx -h \right)^2\right]\right|_{m=m_{_{\!M}}}\right]\\
&+\frac{1}{\D_\kappa}\int_{-\infty}^\infty\frac {\dd x}{\sqrt{2\pi\tilde q}} \left[\left(m_m+\frac{x^2}{2}\right)\ln \left(m_m+\frac{x^2}{2}\right) - \left(m_m+\frac{x^2}{2}\right)\right] \left[\left.\frac{\dd}{\dd m}\exp\left[-\frac{1}{2\tilde q}\left(\frac{x^3}{6} + mx -h \right)^2\right]\right|_{m=m_m}\right]\\
&+\int_{-\infty}^\infty\frac {\dd x}{\sqrt{2\pi\tilde q}}\int_{m_m}^{m_{_{\!M}}}\frac{\dd m}{\D_\kappa} \left(\left(m+\frac{x^2}{2}\right)\ln \left(m+\frac{x^2}{2}\right)-\left(m+\frac{x^2}{2}\right)\right) \frac{\dd^2}{\dd m^2}\exp\left[-\frac{1}{2\tilde q}\left(\frac{x^3}{6} +mx -h \right)^2\right] \ .
\end{split}
\eeq
In summary, we have reduced the computation of the replicon integral $\II_R$ to the sum of perfectly convergent, singularity-free integrals that can be easily evaluated numerically. For the purpose of numerical integration (such as in the case of the algorithms reported above), it is convenient to rescale the integration variable $x$ by $\sqrt{\tilde q}$ and also do the same on $h$:
\beq
x\to x\sqrt{\tilde q} \ \ \ \ \ \ \ h= \Gamma\sqrt{\tilde q}
\eeq
being $\Gamma$ a constant of order one.

\end{widetext}


\end{document}